\documentstyle[preprint,aps]{revtex}
\begin{document}
\draft

\title
{
Step-induced unusual magnetic properties of ultrathin Co/Cu films:
ab initio study
}
\author{
\bf{A. V. Smirnov}$^{(a)}$\cite{adr}
and
 \bf{A. M. Bratkovsky}$^{(b)}$\cite{corr}
}
\vspace{6mm}
\address{$^{(a)}$Institut f\"ur Festk\"orperphysik, Technische Hochschule,
	D-64289 Darmstadt, Germany
}

\address{$^{(b)}$Hewlett-Packard Laboratories, 3500 Deer Creek Road,
		 Bldg. 26U,
		 Palo Alto, CA 94304-1392
}
\date{July 23, 1996}
\maketitle

%\widetext
\begin{abstract}
We have performed ab initio studies to elucidate the unusual magnetic behavior
recently observed in epitaxial Co films upon absorption of
submonolayers of Cu and other materials.
We find that a submonolayer amount of Cu on a stepped Co/Cu (100)  film
changes dramatically the electronic and magnetic structure of the system.
The effect is mainly due to hybridization of Co and Cu $d$-electrons
when copper forms a ``wire'' next to a Co step at the surface.
As a result, a non-collinear 
arrangement of magnetic moments (switching of the easy axis) is promoted.

\end{abstract}	
	
\narrowtext
Recently, a remarkable new phenomenon in magnetism of systems with
reduced dimensionality has been reported. It has been found that 
copper
coverages as small as three-hundredths of a monolayer are sufficient
to rotate discontinuously
the magnetization direction of Co films (up to 20 atomic
layers thick) \cite{weber1}.
The analogous situation takes place also for other overlayers
covered with submonolayers of Ag, Au, and even oxygen
\cite{weber2}. 
Buckley $et$ $al.$\cite{buckley1,buckley2} have also concluded that
in their experiments single 
Cu atom at the surface can affect the behavior of more than 40 Co atoms throughout the 
thickness of the film.
This is perhaps one of the most remarkable new phenomena in ultrathin
film studies, which currently attracts much attention
\cite{weber1,weber2,buckley1,buckley2,schneider,miguel,engel,wulfhekel}
because of potential applicability for magnetic storage devices.
The authors \cite{weber1,weber2,buckley1,buckley2}
have reasonably speculated that the 
observed unusual magnetic behavior can be explained by 
overlayer-induced changes in the 
electronic structure in combination with steps at the surface.
Steps  on the Co film,
which arise from a slight misorientation of the surface,
can act as  sinks for Cu atoms diffusing on it, 
and thus cause Cu ``wires'' to nucleate and grow.

To gain more insight into magnetic behavior of ultrathin films on an atomistic
level, it is necessary to study micromagnetism by {\em ab initio} methods.
These should address
 the effects of small coverages of nonmagnetic 
overlayers as well as surface and interface steps on the magnetic 
properties of such films.
Only a few first-principles calculations of magnetic 
properties of ultrathin Co films have been performed, but only for ideal 
surfaces \cite{wang,zhong}.
Although it is widely recognized that atomic-scale defects can 
strongly affect magnetic properties, to the best of our knowledge,
there have been no previous theoretical studies of the
magnetic nanostructure of ultrathin Co/Cu films with a non-ideal surface, 
although some attempts to calculate magnetic properties with surface 
irregularities  were undertaken (for instance, for Fe/Cr-based 
systems\cite{bluegel2,vega}).

The purpose of the present work
 is  {\em ab initio} analysis 
of the step-induced magnetic properties of thin Co films, including 
the effects of submonolayer quantities of a non-magnetic overlayer,
and their relation to ``magnetic switching''.

For the calculations of the electronic structure, we made use of
the tight-binding (TB) LMTO method \cite{oka84} in real-space (RS) 
\cite{amb_cm}, which we successfully applied previously for studies of 
magnetism in complex disordered systems 
\cite{amb_cm,amb_prb1,amb_prb2,smi_prb}. Here, we use the TB-LMTO method for 
first-principles calculations of both ferromagnetic and constrained
non-collinear magnetic two atomic layers (AL) cobalt films with 
non-ideal surface on the Cu substrate. The specific form of the {\em ab 
initio} TB-LMTO Hamiltonian in the atomic sphere approximation is the 
same as in Ref.~\cite{smi_prb}. Our structural models are based on the
supercell approach: the unit cell contains 144 atomic spheres (some
used to simulate vacuum) positioned at the 
sites of the fcc lattice and distributed in 12 AL for [001] and 
[1$\overline{1}$0] directions and 1 AL for the [110] direction -- Cu atoms of 
the substrate occupy the lower 5 AL, the top 4 AL are filled by the empty
spheres, and the central 3 AL are used for the Co film with 
a step (Fig.\ref{f:fig1}).
The distance between neighbors corresponds to the measured lattice 
constant of fcc-Cu, a$_{\rm Cu}$ = 3.615 \AA, Ref.~\cite{miguel}. 
For the real-space calculations by the recursion method \cite{recurs}
we have used up to $\sim$3000-atom cluster built from the cell according to 
the symmetry of the lattice. We apply periodic boundary conditions in the
plane of the surface. For the perpendicular
Z-direction ([001]), our calculations for 2- and 3-AL cobalt films on a
copper substrate show no noticeable differences from either a
semi-infinite Cu crystal below and vacuum above the Co film
or with  periodicity for  the Z axis;
therefore we shall present our results for the latter case.
It is worth  mentioning that 18 nearest 
neighbors were taken into account for constructing the Hamiltonian 
matrix; all the Cu and empty spheres were equal size, but the Co spheres were
slightly larger ($S_{\rm Co}$/$S_{\rm Cu}$ = $a_{\rm Co}$/$a_{\rm Cu}$ = 1.019) 
\cite{miguel}. In our calculations, we require self-consistent convergence
better than 0.0008 electron/(a.u.)$^3$ for the average root mean square
deviation between the input and output of both electron and spin densities.

To begin with, we discuss the results of our calculations for
``ideal'' reference systems (two atomic layers of Co on
the surface of fcc-Cu and the same with 1 AL (Co or Cu) as overlayer),
 see Table \ref{t:tab1}.
The calculated densities of states (DOS) for
minority spin carriers are shown in
Fig.~\ref{f:fig2}. These states are actively involved in formation of
electronic structure, while majority spin states in Co are almost full.
Our data are in favor of a simple 'bond-counting' argument
that for a pure Co-film  a reduced coordination number 
for atoms in a film narrows the $d$-band and 
increases the exchange splitting. Thus, we observe an enhanced
local magnetic moment at the surface as compared to the bulk crystal 
atoms ($M_{\rm 3AL-Co}/M_{\rm fcc-Co} \sim 1.11 $).
On the other hand, the interaction of Co atoms with a non-magnetic Cu 
overlayer suppresses the magnetic moment in the Co film; we observe a strong 
hybridization of the Cu overlayer $d$-states with
corresponding Co minority $d$-states (Fig.\ref{f:fig2}).
The same tendency has been noted in Ref.~\cite{zhong} for a 1 AL Co film.
One can clearly see that Co DOS shows strong changes in the energy
range down to about 0.15 Ry below the Fermi level. The states involved
are mainly Co $3z^2-1$, $xz$, and $yz$, which change abruptly upon
deposition of Cu \cite{zhong}.

Now we are in a position to analyze possible changes of magnetic 
properties due to the surface steps. We consider the
 case of 2.5 AL Co films (Fig.\ref{f:fig1}),
where there is a Co terrace of equal size with a trough (both infinite in 
the [110] direction and replicated in the [1$\bar{1}$0] direction).
Such a surface defect results in a substantial increase of magnetic 
moments on the Co surface atoms in the trough and some decrease of 
the moments on other
atoms (in comparison with Co atomic layers of the reference system). 
Near the steps,  a discontinuity (kink) in the value of
the magnetic moment is clearly seen.
The average magnetic moment on Co atoms is about 1.70~$\mu_B$/Co,
the same as for the 3 AL Co film.  Average values for each layer are given
in Table \ref{t:tab1}. 
Upon deposition of Cu wires one would expect strong changes in magnetic
behavior of the system due to mainly
two reasons:  Co-Cu hybridization and electron charge redistribution.
Incidentally,
the fact that the majority Co d-band is almost full means that charge
transfer would change mainly a number of minority carriers on a site, so
that the number of transferred electrons will be approximately equal to the
change of a magnetic moment on the same site.

Results of calculations for 2.5 AL Co films with 1-, 2- and 3-Cu wires
placed in the trough near the step are presented in 
Table \ref{t:tab1} and Fig.~\ref{f:fig3}.
The deposited Cu substitutes an
empty site adjacent to a step and consequently sqeezes out about
0.6~electron. This charge redistribution involves all (!)
terrace Co sites causing significant changes in magnetic moments.
Namely,
addition of {\em just one} overlayer chain of Cu atoms
{\em increases} the terrase magnetic moment on $0.06~\mu_B/atom$ and
{\em decreases} the moment on the nearest neighboring Co sites up to
11\%; the moment for the Co atoms in the second plane
beneath the copper wire increases from 1.56 $\mu_B$ up to 1.64 $\mu_B$.

There are many ways to add a second Cu wire.
We have considered only the one where Cu atoms are adsorbed on the
terrace level (in the 3rd ``central'' AL). Our calculations show that
the growth of a continuous Cu overlayer is preferable, that is,
the energy of systems with 2 adjacent Cu chains attached to the Co steps is 
lower than for the case when each Cu wire is placed symmetrically
near separate atomic steps.

The dependence of the nanomagnetic structure on the number of Cu overlayer
atoms is shown on Fig.\ref{f:fig3}. The values of the Co magnetic moments
under the copper steadily decrease when the amount of Cu increases
(the average magnetic moment on Co shows the same tendency,
Table \ref{t:tab1}),
approaching its mean value (1.54 $\mu_B$) for a completed Cu overlayer.
At the same time, the  magnetic moments on Co terrace atoms go up,
and, therefore, the jump in the moment value increases together with the
number of deposited Cu atoms.
This, as well as the charge redistribution, may promote a frustration of the
directions of magnetic moments, as we shall now discuss.

Because of the low symmetry of the system we consider, there are no constraints 
preventing the system from developing a complex (non-collinear) magnetic order.
To investigate this possibility, we have performed a series of calculations
with a `frozen' magnetic configuration as shown in
Fig.\ref{f:fig4}. There the angle between the moments on two cobalt wires close to 
the Cu/Co boundary
and the moments on all other atoms has been fixed at a right angle
with respect to each other. 
We have then performed self-consistent calculations 
for the charge and spin densities (the procedure is analogous to
that used in a ferromagnetic case \cite{constraint}). 
We have found that this constrained {\em non-collinear} configuration is
slightly {\em lower} in energy ($< 2$ mRy/atom) than the {\em collinear}
ferromagnetic one\cite{constraint}. This clearly shows a tendency
towards a non-collinear magnetic order, which has been also
confirmed by  calculations with unrestricted spin directions.

%\section{Conclusions}

Summarizing, for the first time we have performed {\em ab initio} 
studies of ultrathin 
Co/Cu films with consideration of the effects of both a stepped surface
and Cu overlayer atoms on the magnetic properties. We have found that they
both can radically change the nanomagnetic
structure of the film. 
We have found that deposition of a copper `wire' causes a long-range
charge and moment redistribution over a large number of adjacent Co sites,
as a result of strong $d$-$d$ hybridization between Co and Cu atoms.
The results of calculations with a constrained magnetic
arrangement have clearly showed a tendency towards a non-collinear magnetic
ordering in the ultrathin Co films with a non-ideal surface due to
a hybridization between $d$-electrons on Cu and Co.
% even in the absence  of ordinary mechanisms of magnetic anisotropy such
%as spin-orbit coupling and magnetoelastic effects \cite{gay}).

%\acknowledgments

The authors are indebted to J.~K\"ubler,
 L.~Sandratskii, and R.S.~Williams for helpful discussions.
AVS has been supported
by the Alexander von Humboldt Foundation.

\begin{table}   %%%%  TABLE I %%%%%%%%%%%
\caption{ The calculated values of the magnetic moments for 
various Co films (in $\mu_B$), and for different numbers of Cu 
overlayer atoms (for details see text).
For films with surface irregularities, the averaged
values for each atomic layer (AL) are given. $<M>$ is the average 
magnetic moment on all Co atoms. 
For comparison, the measured value for a thick fcc Co
film (on Cu) is  1.68 $\mu_B$ [19],
and the calculated magnetic 
moment for fcc Co crystal, M$_{\rm fcc-Co}$ is
1.65~$\mu_B$ (1.66$\mu_B$ from the LMTO 
calculations [19] ).
}
\begin{tabular}{ccccc}
Film  & 1st AL   & 2nd AL   & 3rd AL   &  $<M>/\mu_B$\\
\hline
\tableline
2AL-Co & 1.55 & 1.83 & 0.03 (ES)&  1.69 \\
3AL-Co & 1.66 & 1.60 & 1.83&  1.70 \\
\tableline
2.5AL-Co&1.603&1.755&1.771&1.697 \\ 
2.5AL-Co$+$1 Cu-wire&1.607&1.733&1.829&1.702 \\
2.5AL-Co$+$2 Cu-wires&1.613&1.702&1.842&1.694 \\ 
2.5AL-Co$+$3 Cu-wires&1.619&1.668&1.850&1.685 \\
\tableline
2AL-Co$+$1AL-Cu & 1.64 & 1.54 & 0.01 (Cu)& 1.59 \\
\end{tabular}
\label{t:tab1}
\end{table}

%%%%%%%%%%%%%%%%%%%%%%%%%%%%%%%%%%%%%%%%%%%%%%%%%%%%%%%%%%%%%%%%%%%%%%%%
%%%%%%%%%%%%%%%%%%%%%%%%%%%%%%%%%%%%%%%%%%%%%%%%%%%%%%%%%%%%%%%%%%%%%%%%
%
%        F I G U R E S
%
%%%%%%%%%%%%%%%%%%%%%%%%%%%%%%%%%%%%%%%%%%%%%%%%%%%%%%%%%%%%%%%%%%%%%%%%
%%%%%%%%%%%%%%%%%%%%%%%%%%%%%%%%%%%%%%%%%%%%%%%%%%%%%%%%%%%%%%%%%%%%%%%%

\begin{figure} %%%%   FIGURE   1   %%%%
\caption{Structural model and magnetic nanostructure for
2.5 AL Co films on a Cu substrate (schematic). Co atoms are indicated by
a bold contour. Vacuum is simulated by the empty spheres (ES).
The values of the
 magnetic moment are shown inside each site. All magnetic
moments are parallel to each other.
In the present calculations all atoms of the ``central'' layers are assumed
to be
non-equivalent; for other sites non-equivalence is considered only for 
different atomic layers, excluding boundary layers.
}
\label{f:fig1}
\end{figure}

\begin{figure} %%%%   FIGURE   2   %%%%
\caption{The local density of states (DOS) for spin-{\em down}
(minority) 
$d$ states of 2 AL
Co films with Cu (a) and Co (b) mono-overlayers; dashed lines
correspond to the DOS for fcc Cu (a) and Co (b) bulk crystals.
}

\label{f:fig2}
\end{figure}

\begin{figure} %%%%   FIGURE   3   %%%%
\caption{Magnetic moments for 2.5 AL Co films with
one (a), two (b), and three (c) Cu wires. The position of the Cu atoms
is shown by dashed lines. The Co atoms with numbers 2 and 24 correspond
to the atoms at the terrace middle. The structural model is shown in Fig.1.
}
\label{f:fig3}
\end{figure}         

\begin{figure} %%%%   FIGURE   4   %%%%
\caption{Structural model and local magnetic moments for
2.5 AL Co films  on a Cu substrate with 2 Cu wires near the atomic
step (schematic).
For symbols, see Fig.1. The magnetic moments of the 
unshadowed Co atoms are
perpendicular to the rest of the Co atoms.
}
\label{f:fig4}
\end{figure} 


\begin{references}

\bibitem[*]{adr} Permanent address: Kurchatov Institute, 123182 Moscow, Russia.
\bibitem[+]{corr} Corresponding author. Electronic address: alexb@hpl.hp.com

\bibitem{weber1} W. Weber, C. H. Back, A. Bischof, D. Pescia, and 
R. Allenspach, Nature {\bf 374}, 788 (1995). 
\bibitem{weber2} W. Weber, C. H. Back, U. Ramsperger,
A. Vaterlaus, and R. Allenspach, Phys. Rev. B{\bf 52}, R14400 (1995).
\bibitem{buckley1} M. E. Buckley, F. O. Schumann, and J. A. C. Bland, 
\prb {\bf 52}, 6596 (1995).
\bibitem{buckley2} M. E. Buckley, F. O. Schumann, and J. A. C. Bland,
J. Phys. Condens. Matt. {\bf 8}, L147 (1996).

\bibitem{schneider} C. M. Schneider {\em et al},
%, P. Bressler, P. Schuster, J. Kirschner, J. J. de Miguel, and R. Miranda, 
Phys. Rev. Lett. {\bf 64}, 1059 (1990).
\bibitem{miguel} J. J. de Miguel {\em et al}, J. Magn. Magn. Mater.
{\bf 93}, 1 (1991).
\bibitem{engel} B. N. Engel, M. H. Wiedmann, R. A. Van Leeuwen, 
and C. M. Falco, J. Appl. Phys. {\bf 73}, 6192 (1993).
\bibitem{wulfhekel} W. Wulfhekel  {\em et al}, Phys. Rev. B{\bf 50},
16074 (1994).  
\bibitem{wang} D. S. Wang, R. Wu, and A. J. Freeman, 
J. Magn. Magn. Mater.{\bf 129}, 237 (1994).
\bibitem{zhong} L. Zhong, M. Kim, X. Wang, and A. J. Freeman,
Phys. Rev. B{\bf 53}, 9770 (1996).
\bibitem{bluegel2} S. Bl\"ugel, D. Pescia, and P. H. Dederichs, Phys. 
Rev. B{\bf 39}, 1392 (1989).
\bibitem{vega} A. Vega, C. Demangeat, H. Dreyss\'e, and A. Chouairi,
Phys. Rev. B{\bf 51}, 11546 (1995); A. Vega, D. Stoeffler, H. Dreyss\'e,
and C. Demangeat, Europhys. Lett. {\bf 31}, 561 (1995).
\bibitem{oka84} O. K. Andersen and O. Jepsen, Phys. Rev. Lett.
{\bf 53}, 2571 (1984).
\bibitem{amb_cm} A. M. Bratkovsky and A. V. Smirnov, J. Phys. Cond.
Matt. {\bf 5}, 3203 (1993).
\bibitem{amb_prb1} A. M. Bratkovsky and A. V. Smirnov, Phys. Rev. 
B{\bf 48}, 9606 (1993).
\bibitem{amb_prb2} A. M. Bratkovsky, A. V. Smirnov, D. Nguyen Manh,
and A. Pasturel, Phys. Rev. B{\bf 52}, 3056 (1995).
\bibitem{smi_prb} A. V. Smirnov and A. M. Bratkovsky, 
\prb {\bf 53}, 8515 (1996); Europhys. Lett. {\bf 33}, 527 (1996); 

\bibitem{recurs}  R. Haydock, Sol. State Phys. {\bf 35}, 215 (1980).
\bibitem{liu_co} X. Liu {\em et al.}, Phys. Rev. B{\bf 53}, 12166 (1996).
\bibitem{constraint} 
With regards to the constrained calculations it  is worth mentioning that,
according to Ref.~\cite{ls_kuv}, for bcc-Fe a
calculation with a fixed magnetic configuration (as in present case) gives 
results almost equivalent to those where the constraint has been imposed by
applying  an external magnetic field (in more rigorous approach). However,
in the case of fcc-Ni the difference between the two methods is noticeable.
\bibitem{ls_kuv} L. M. Sandratskii and E. N. Kuvaldin, J. Phys. Cond. 
Matt. {\bf 3}, 7663 (1991).


\end{references}
\end{document}